\documentclass[nobibnotes,preprintnumbers,aps,prl,onecolumn,superscriptaddress,showpacs,sort&compress]{revtex4}				 %

\usepackage{graphicx,epstopdf,hyperref}		

\usepackage{titlesec}																											     %
\begin{document}																													 %
\vspace*{-0.3cm}																													 %

\title{Exact Diagonalization Study of an Extended Hubbard Model \\for a Cubic Cluster at Quarter Filling}

\author{K. Sza\l{}owski}
\thanks{corresponding author; e-mail: kszalowski@uni.lodz.pl}
\affiliation{Department of Solid State Physics, Faculty of Physics and Applied Informatics, University of \L{}od\'z, ulica Pomorska 149/153, PL 90-236 \L{}od\'z, Poland}
\author{T. Balcerzak}
\affiliation{Department of Solid State Physics, Faculty of Physics and Applied Informatics, University of \L{}od\'z, ulica Pomorska 149/153, PL 90-236 \L{}od\'z, Poland}
\author{M. Ja\v{s}\v{c}ur}
\affiliation{Department of Theoretical Physics and Astrophysics, Faculty of Sciences, P. J. \v{S}af\'{a}rik University, Park Angelinum 9, 041 54 Ko\v{s}ice, Slovakia}
\author{A. Bob\'{a}k}
\affiliation{Department of Theoretical Physics and Astrophysics, Faculty of Sciences, P. J. \v{S}af\'{a}rik University, Park Angelinum 9, 041 54 Ko\v{s}ice, Slovakia}
\author{M. \v{Z}ukovi\v{c}}
\affiliation{Department of Theoretical Physics and Astrophysics, Faculty of Sciences, P. J. \v{S}af\'{a}rik University, Park Angelinum 9, 041 54 Ko\v{s}ice, Slovakia}

\begin{abstract}
In the paper the thermodynamics of a cubic cluster with 8 sites at quarter filling is characterized by means of exact diagonalization technique. Particular emphasis is put on the behaviour of such response functions as specific heat and magnetic susceptibility. The system is modelled with extended Hubbard model which includes electron hopping between both first and second nearest neighbours as well as coulombic interactions, both on-site and between nearest-neighbour sites. The importance of hopping between second nearest neighbours and coulombic interactions between nearest neighbours for the temperature dependences of thermodynamic response functions is analysed. In particular, the predictions of Schottky model are compared with the calculations based on the full energy spectrum.
\end{abstract}


\maketitle

%
\section{Introduction, model and method}
The Hubbard model and its extensions, being fundamental models for strongly correlated systems, still constitute a challenge and their thermodynamics attracts considerable attention \cite{Duffy,Noce,Becca,Paiva,Paiva2,Harir,Lopez,Nath,Nath2}. The exact thermodynamic solutions are known only for a limited range of systems, including zero-dimensional ones, for which exact diagonalization can be performed \cite{ref1,ref2,ref3,ref4,ref5,ref6}, albeit this approach requires significant computational resources. 
In the present study we deal with a cubic, zero-dimensional cluster consisting of 8 atoms (sites) filled with 4 electrons (what constitutes quarter-filling case). It is described by the following Hamiltonian of extended Hubbard model:
\begin{eqnarray}
\mathcal{H}&=&\!\!-t_1\!\!\sum_{\left\langle i,j\right\rangle,\sigma}^{}{\!\left(c^{\dagger}_{i,\sigma}c_{j,\sigma}+h.c.\right)}-t_2\!\!\!\!\sum_{\!\left\langle\left\langle i,j\right\rangle\right\rangle,\sigma}^{}{\!\!\left(c^{\dagger}_{i,\sigma}c_{j,\sigma}+h.c.\right)}\nonumber\\&&+\,U\,\sum_{i}^{}{n_{i,\uparrow}n_{i,\downarrow}}+V\!\!\!\!\sum_{\left\langle i,j\right\rangle,\sigma,\sigma'}^{}{\!\!n_{i,\sigma}n_{j,\sigma'}}.
\end{eqnarray}
Here, $t_1$ and $t_2$ denote the hopping integrals between nearest-neighbours (NN) and second NN, respectively, while $U$ is the on-site coulombic interaction energy and $V$ is the energy of coulombic interaction between electrons at NN sites. The operators $c^{\dagger}_{i,\sigma}$ ($c_{i,\sigma}$) create (annihilate) the electron with spin $\sigma=\uparrow,\downarrow$ at site $i$ and $n_{i,\sigma}=c^{\dagger}_{i,\sigma}c_{i,\sigma}$. In order to solve our model, we exploit the exact diagonalization approach, which we have already developed in \cite{ref1} for analogous cluster. The eigenvalues and eigenvectors of the Hamiltonian matrix are calculated with \textit{Wolfram Mathematica} software \cite{ref7}. Further thermodynamic analysis is performed within canonical ensemble formalism \cite{ref1}, which is based on determination of statistical sum for the system, from which all further thermodynamic quantities can be derived \cite{ref1}. In the present study we focus our interest on specific heat and magnetic susceptibility of the system. 

It has been established in our earlier study \cite{ref1} that the temperature dependence of specific heat exhibits double-peak structure, while the analogous dependence of magnetic susceptibility shows a single peak. The sensitivity of that maxima to the value of on-site coulombic energy $U$ was discusses and a good applicability of Schottky model was found. The aim of the present study is to analyse the importance of $t_2$ and $V$ (which parameters extend the usual Hubbard model) on the behaviour of thermodynamic response functions such as specific heat and magnetic susceptibility. The Schottky model, being a useful tool for understanding the thermodynamics of zero-dimensional systems \cite{ref1,ref8}, is also worthy of investigation in that context.

\section{Numerical results and discussion}
\begin{figure}[h!]
\includegraphics[width=0.95\columnwidth]{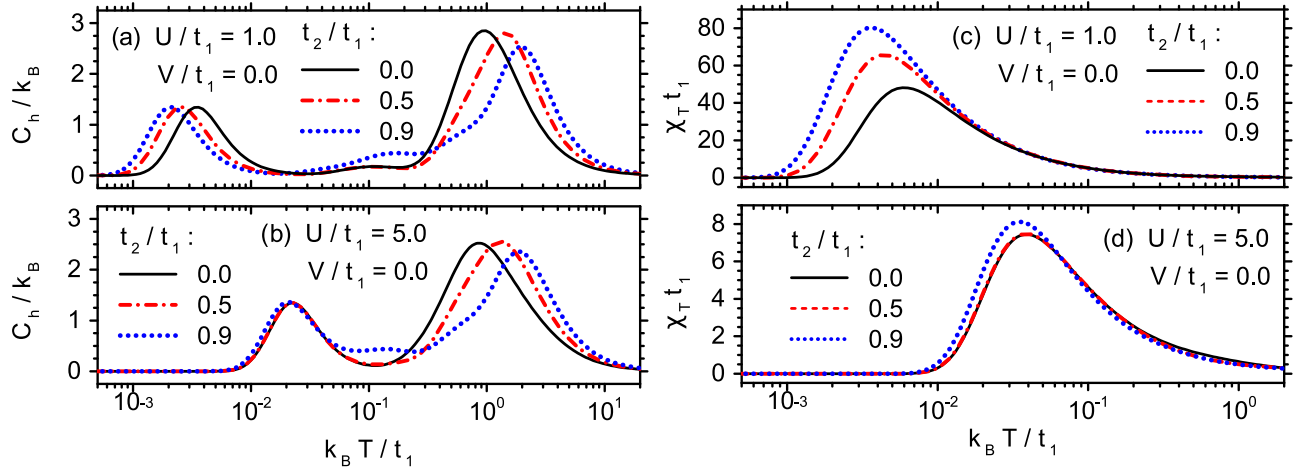}
\caption{Temperature dependence of specific heat (left) and magnetic susceptibility (right) for various ratios of hopping between second NN and NN, for $U / t_1 =$ 1.0 and 5.0.}
\label{fig1}
\end{figure}

Let us commence the analysis from the influence of the hopping between second NN on the thermodynamic parameters. We consider the range of $0 < t_2 < t_1$, which seems physically justified. The effect of second NN hopping on the specific heat can be followed in Fig.~\ref{fig1}(a,b). As it is visible in Fig.~\ref{fig1}(a), for $U / t_1  = 1.0$, the presence of $t_2$ affects the positions of both low- and high-temperature maximum of the specific heat. The first one tends to shift towards lower temperatures, while the second one exhibits an opposite tendency. Both shifts are quite significant (please note the logarithmic scale). For very strong $t_2$ a third, intermediate maximum tends to build up, but with rather reduced height. On the contrary, for the case of $U / t_1  = 5.0$, illustrated in Fig.~\ref{fig1}(b), only the high-temperature peak of specific heat is sensitive to $t_2$, while the low-temperature maximum shows no tendency to shift and the intermediate maximum appears. 
The behaviour of magnetic susceptibility is illustrated in Fig.~\ref{fig1}(c,d). The single maximum tends to shift towards lower temperatures when $t_2$ increases only for the lower value of $U / t_1  = 1.0$, whereas it does not move for $U / t_1  = 5.0$. Therefore, the behaviour of susceptibility maximum follows the trend for the low-temperature peak of the specific heat.
\begin{figure}[h!]
\includegraphics[width=0.95\columnwidth]{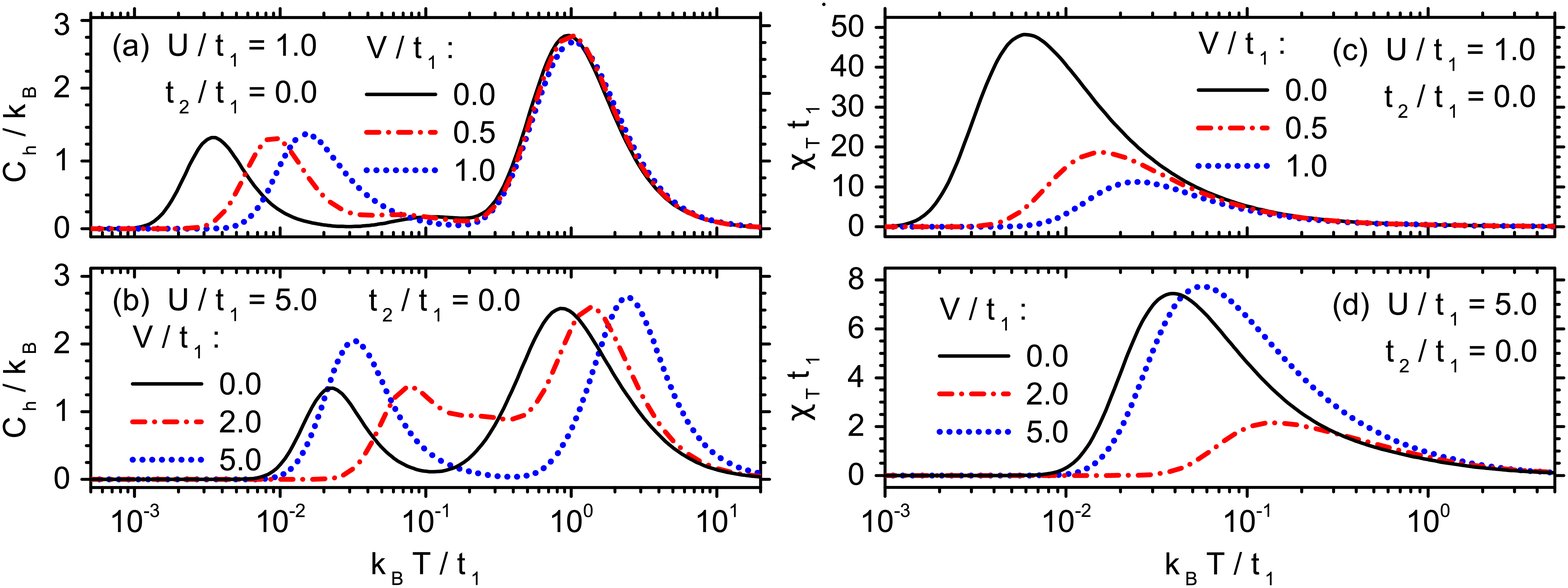}
\caption{Temperature dependence of specific heat (left) and magnetic susceptibility (right) for various energies of coulombic interaction between NN, for $U / t_1 =$ 1.0 and 5.0.}
\label{fig2}
\end{figure}
The influence of coulombic interaction between electrons located at NN sites is again studied for the physically relevant range of $0<V<U$. The effect of introducing $V>0$ can be followed in Fig.~\ref{fig2}(a,b) for specific heat. When $U / t_1  = 1.0$ (Fig.~\ref{fig2}(a)), the low-temperature peak exhibits high mobility and becomes shifted towards higher temperatures, by approximately order of magnitude. On the contrary, the position of the second, high-temperature maximum remains untouched. Both extrema also tend to conserve their heights. The behaviour of specific heat for $U / t_1  = 5.0$, as shown in Fig.~\ref{fig2}(b) is more complex. The high-temperature peak is strongly shifter towards higher temperatures. On the contrary, the position of low-temperature maximum shows a non-monotonic dependence on $V / t_1$, with initial increase and further return to lower values of $T_{max}$. Moreover, the height of this extremum also varies, since for higher V it becomes more pronounced. At the same time, at moderate values of $V / t_1$, the specific heat at intermediate temperatures between the peaks is also significantly elevated, which effect vanishes when $V$ further increases.

The evolution of the temperature dependence of magnetic susceptibility with varying $V$ is shown in Fig.~\ref{fig2}(c,d). For $U / t_1  = 1.0$ (Fig.~\ref{fig2}(c)) the maximum shifts very significantly towards higher temperatures when $V$ is switched on and simultaneously its height is greatly reduced. The situation is changed when $U$ is stronger, i.e. for $U / t_1  = 5.0$, as seen in Fig.~\ref{fig2}(d). There, the position of maximum depends non-monotonically on $V$ (similarly to the dependence of low-temperature maximum of specific heat, with initial increase and further decrease of $T_{max}$). Moreover, the value of susceptibility at extremum is reduced for moderate $V / t_1$ and it rises back when $V$ becomes stronger.
\begin{figure}[h!]
\includegraphics[width=0.6\columnwidth]{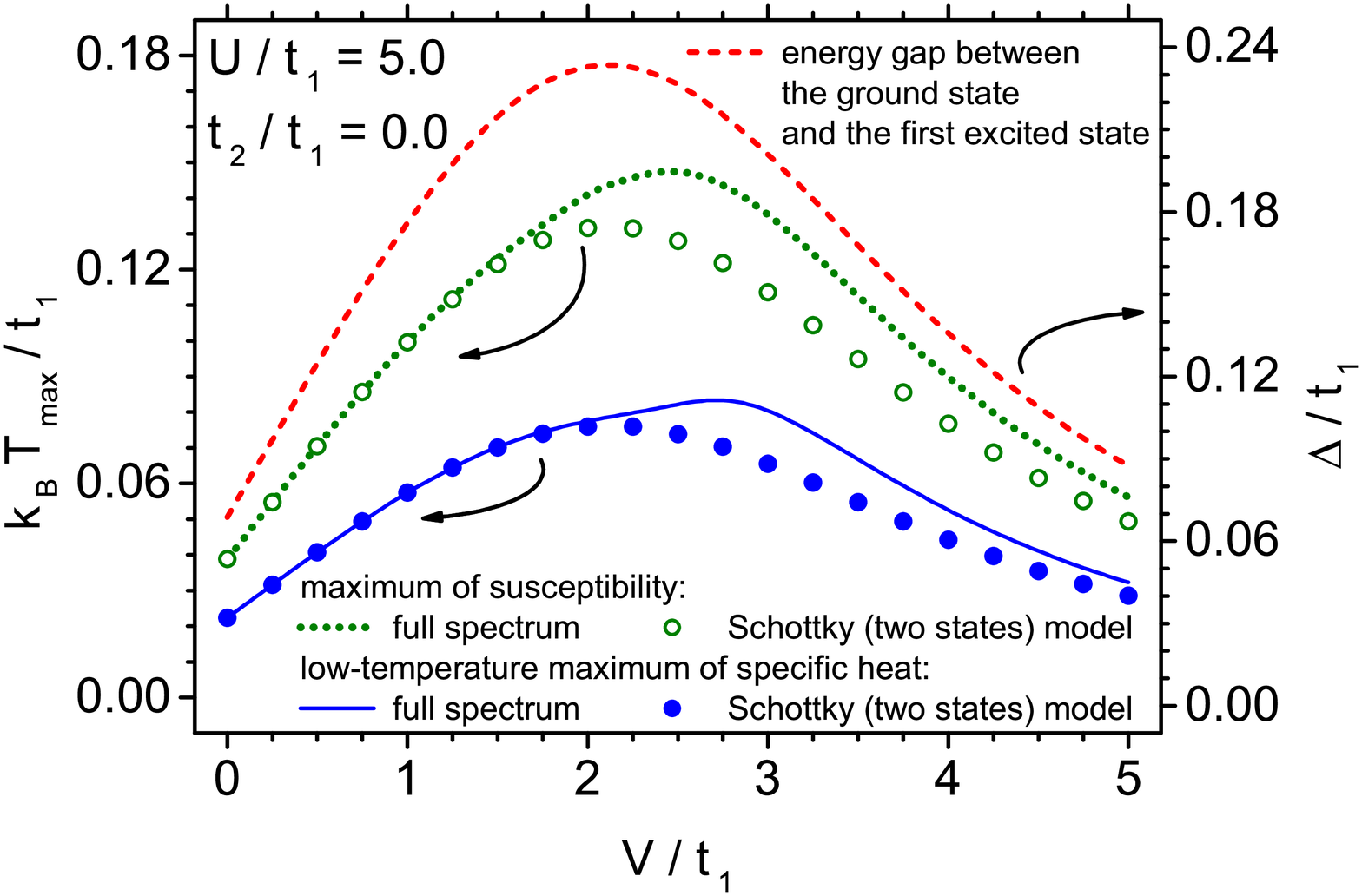}
\caption{The position of the low-temperature maximum of specific heat and the maximum of magnetic susceptibility together with the predictions of Schottky model (left vertical scale) and the energy gap between ground state and first excited state (right vertical scale) as a function of coulombic interactions between NN, for $U / t_1 =$ 5.0.}
\label{fig3}
\end{figure}
\begin{figure}[h!]
\includegraphics[width=0.6\columnwidth]{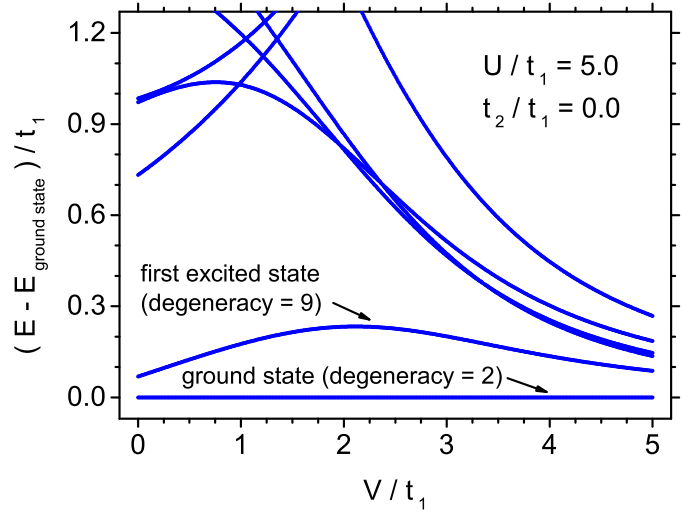}
\caption{The energies of the states lying close to the ground state as a function of the coulombic interactions between NN, for $U / t_1 =$ 5.0. The energy of the ground state is set to zero. }
\label{fig4}
\end{figure}
Our previous study \cite{ref1} revealed that the low-temperature maximum of specific heat and the maximum of susceptibility can be explained by referring to the Schottky model, involving a ground state and a first excited state, separated with an energy gap $\Delta$. The relation between the energy gap and the positions of both extrema $T_{max}$ was derived in our work \cite{ref1} for our system of interest. 

The detailed dependence of the position of the mentioned extrema $T_{max}$ on $V$ for $U / t_1 = 5.0$ can be followed in Fig.~\ref{fig3}. Such a choice corresponds to Fig.~\ref{fig2}, where a non-monotonic behaviour as a function of $V$ was seen. This behaviour is confirmed in Fig.~\ref{fig3}, where the temperature at which the extremum is reached achieves largest values around $2.4\lesssim V / t_1 \lesssim 2.6$. For comparison, the normalized energy gap between the ground state and the first excited state is plotted in the same figure. The Schottky model predicts that the temperature $T_{max}$ should be proportional to $\Delta$ (see \cite{ref1}) and the predictions of $T_{max}$ based on \cite{ref1} for specific heat maximum and susceptibility maximum are also indicated in Fig.~\ref{fig3} (with circles). It is visible that for $V / t_1 < 2$ the consistency between the Schottky model and the calculations involving the full energy spectrum of the system is very good. However, for stronger $V$ a discrepancy arises and the maximum of $T_{max}$ with respect to $V / t_1$ is reached at higher $V$ that the Schottky model shows. This kind of behaviour can be explained on the basis of Fig.~\ref{fig4}, which shows the dependence of energies of a few states lying close to the ground state on $V$. For low $V$, the separation in energy between ground state and the first excited state is much smaller than the energy difference between first and second excited state. Therefore, the conditions for Schottky approximation are well fulfilled. When $V$ rises, the second excited state gets closer to the first one and its contribution grows, yielding the discrepancy between the predictions of Schottky model and exact calculations.
\begin{figure}[h!]
\includegraphics[width=0.6\columnwidth]{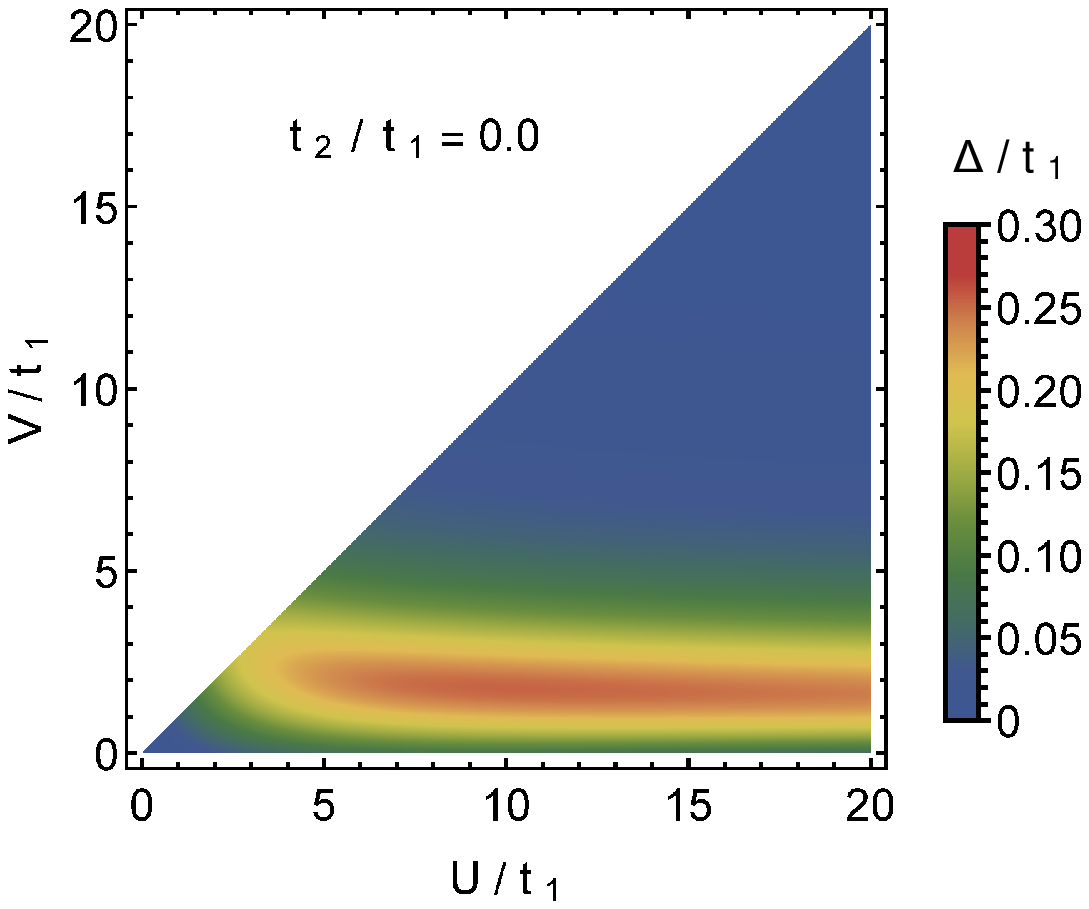}
\caption{The energy gap between the ground state and the first excited state as a function of the energy of on-site coulombic interactions and coulombic interactions between NN. }
\label{fig5}
\end{figure}
The evolution of the energy gap as a function of $U/t_1$ and $V<U$  can be followed in Fig.~\ref{fig5}. It is evident that the gap tends to reach a maximum value at some low, but non-zero $V/t_1$, which is almost independent on $U/t_1$. Moreover, the maximum gap value at this $V/t_1$ is weakly sensitive to $U/t_1$ unless it is very low. Therefore, the behaviour illustrated for $U/t_1=5.0$ is representative also to higher values of $U$.

\section{Conclusions}
The study revealed the sensitivity of peaks of specific heat and magnetic susceptibility to such extended Hubbard model parameters as $t_2$ and $V$ at quarter filling for a cubic cluster. The magnitude of the influence depends vitally on the value of $U$. The conditions for applicability of Schottky model were established. Futher studies of clusters with other geometry are well motivated.
\section{Acknowledgement}
This work has been supported by Polish Ministry of Science and Higher Education on a special purpose grant to fund the research and development activities and tasks associated with them, serving the development of young scientists and doctoral students.

\end{document}